\documentclass[%
 reprint,
amsmath,amssymb,
aps,
]{revtex4-2}

\usepackage{graphicx}
\usepackage[T1,T2A]{fontenc}
\usepackage[utf8]{inputenc}
\usepackage[russian]{babel}
\usepackage{amsmath}
\usepackage{lineno}
\usepackage{tikz}
\usepackage{multirow}
\usepackage{caption}
\usepackage{subcaption}
\usepackage{dcolumn}
\usepackage{bm}

\usepackage{pgfplots}
\pgfplotsset{compat=1.15}
\usepackage{mathrsfs}
\usetikzlibrary{shapes.geometric, arrows}

\tikzstyle{startstop} = [rectangle, rounded corners, minimum width=1.8cm, minimum height=1cm,text centered, draw=black, fill=red!30]

\tikzstyle{arrow} = [thin,->,>=stealth]

\begin{document}

\preprint{APS/123-QED}

\title{Influence of the finite transverse size of the accelerating region on the relativistic feedback}

 \author{Alexander Sedelnikov}
 \altaffiliation{Moscow Institute of Physics and Technology, Moscow, 117303, Russian Federation; Lebedev Physical Institute RAS}
\email{sedelnikov.as@phystech.edu}

\author{Egor Stadnichuk}
\altaffiliation{Moscow Institute of Physics and Technology, Moscow, 117303, Russian Federation;
 HSE University, Moscow 101000 Russia}
 \email{yegor.stadnichuk@phystech.edu}

\author{Eduard Kim}
  \altaffiliation{Moscow Institute of Physics and Technology, Moscow, 117303, Russian Federation;
   Institute for Nuclear Research of RAS, Moscow 117312}
 \email{kim.e@phystech.edu}

\author{Oraz Anuaruly}
  \altaffiliation{Moscow Institute of Physics and Technology, Moscow, 
 Lebedev Physical Institute RAS}
 \email{orazanuaruly@gmail.com}

 \author{Daria Zemlianskaya}
  \altaffiliation{Moscow Institute of Physics and Technology, Moscow, 117303, Russian Federation;
 Institute for Nuclear Research of RAS, Moscow 117312}
\email{zemlianskay.d@phystech.edu}

 \affiliation{
}




\date{May 29, 2023}

\begin{abstract}
Terrestrial gamma-ray flashes (TGFs) are commonly associated with relativistic runaway electron avalanches (RREAs). However, research shows that a single RREA cannot generate observable TGF fluxes. In an attempt to settle this issue the relativistic feedback mechanism was suggested by Joseph Dwyer. The Monte Carlo simulations and analytical descriptions of this type of feedback assume that acceleration region has a large size in a plane perpendicular to the direction of the electric field. Therefore these studies do not take into account transverse diffusion of RREAs starting points and the finite transverse size of the accelerating region. Electrons created by the feedback outside this region can not be accelerated by the electric field and form an avalanche, which may lead to a decrease in the total number of new avalanches and an increase in the requirements for self-sustaining RREA production by the feedback. In this article the transverse propagation of avalanches starting points was described using a modified two-dimensional diffusion equation. A correction to the criterion for self-sustaining production of RREAs was obtained. Monte Carlo simulation was also performed to calculate the correction for the feedback coefficient. 
\end{abstract}

\maketitle


\section{Keypoints}\label{keypoints}
\begin{itemize}
\item The influence of a finite transverse size of the accelerating region on RREAs dynamics was analytically considered.
\item Taking diffusion into account does not make a significant contribution to the feedback coefficient when transverse size of accelerating region much larger than its longitudinal one.
\item For a transverse size comparable to the longitudinal one, diffusion leads to a significant decrease in the feedback coefficient and a reduction in the number of avalanches in new generations.
\end{itemize}

\renewcommand{\figurename}{Fig.}

\section{\label{intro}Introduction}

One of the unsolved problems in atmospheric physics is the construction of a model of Terrestrial Gamma-ray Flashes (TGFs). This phenomenon was first discovered in 1994 by the Compton Gamma Ray Observatory \cite{doi:10.1126/science.264.5163.1313} and was observed by other space gamma-ray observatories such as Fermi \cite{mailyan2016spectroscopy}, which were created for observing gamma radiation from astrophysical sources.  It has been established that avalanches of relativistic runaway electron avalanches (RREAs) accelerated by an electric field in thunderclouds might be the sources of these flashes \cite{dwyer2012high}. 

The force acting on relativistic electrons from the accelerating field may exceed losses in interactions with air molecules \cite{wilson1925acceleration}.  Such electrons are called runaway electrons. They produce new runaway electrons, leading to the formation of an avalanche \cite{gurevich1992runaway,Gurevich1992}. The dynamics of avalanches is significantly influenced by feedback mechanisms studied by Joseph Dwyer \cite{Dwyer2003}.  As a result of feedback,  the number of electrons is growing and new avalanches can be created.  

There are positron and gamma feedback mechanisms.  Positron feedback can be described as follows.  An avalanche of runaway electrons radiates gamma-rays. These gamma-rays are generated by electron-positron pairs and positrons begin to propagate in the direction opposite to the direction of the electron avalanche. Then the positrons ionize the air at the beginning of the region, which leads to the formation of new RREA. The gamma feedback mechanism is based on the fact that radiated gamma rays are scattered backward and then, at the beginning of the accelerating region,  generate new runaway electrons via Compton scattering or the photoelectric effect.  Number of avalanches in new generation divided by number of avalanches in previous generation is called feedback coefficient.  In other words, it is a probability that the RREA reproduces itself through relativistic feedback. If this  coefficient is greater than one,  avalanches multiplication becomes self-sustainable.  This regime is called infinite feedback because if the electric field strength is constant, the process will never end, and the number of relativistic particles will be unlimited. It is extremely important to understand the conditions under which this regime occurs, because exactly infinite feedback greatly increases the number of runaway electrons, and therefore, can be used to describe the high flux of photons in TGF. \cite{sarria2021constraining, https://doi.org/10.1029/2007JD009248}.

For relatively low electric field strength, positron feedback dominates over gamma-ray feedback \cite{dwyer2007relativistic}, which motivates to study the positron feedback mechanism in the first place.  The criterion of infinite positron feedback was derived in the paper \cite{stadnichuk2022criterion}. This work does not take into account diffusion of RREAs and the finite transverse size of the accelerating region. RREAs of new generations resulting from feedback may be created outside the acceleration region, which may lead to a decrease in the number of new avalanches and an increase in the requirements for self-sustaining RREA production by the feedback.

Balloon measurements showed that there are regions in the thundercloud where electric field exceeds threshold field (the minimum field required for the formation of runaway electrons) \cite{marshall2005observed, https://doi.org/10.1029/95JD00020}. However, in view of the peculiarities of balloon measurements, it is difficult to draw conclusions about the transverse size of the overthreshold regions, which may affect infinite feedback. Therefore, it is necessary to estimate the dependence of the criterion of the infinite feedback regime on transverse size. To settle this issue in this article, an correction to the feedback coefficient was derived. Furthermore, using Geant4 simulation, its value was estimated.

\section{Avalanches diffusion}\label{sec2}

To describe the diffusion of RREAs via relativistic feedback,  a simple diffusion equation can be used with an additional term responsible for the multiplication of avalanches. We will consider an accelerating region with a uniform electric field directed along the z-axis. The avalanche coordinate will be associated with the coordinate of the primary electron from which it was formed, and for simplicity only the two-dimensional distribution of the avalanche will be considered, without taking into account the z coordinate of the beginning of the avalanche. Therefore , the equation describing the concentration of avalanches is
\begin{equation}\label{eq3.1}
     \frac{\partial n_a}{\partial t} -  \triangledown \cdot (D \cdot \triangledown n_a) - \frac{n_a}{\tau^*} = n_s
\end{equation}

where $n_a$ is a two-dimensional distribution of the RREA starting points. $n_s$ is a source function , and $\tau^*=\frac{\tau}{ln \Gamma}$. The last term describes an increase in the number of avalanches due to the feedback factor $\Gamma$ during the time of formation of a new generation $\tau$.
For only one initial avalanche in the center of coordinate system $n_s(x,y,t) = \delta (x) \delta (y) \delta (t)$
the solution of equation (\ref{eq3.1}) will be Green's function, which in the polar coordinate system is:

\begin{equation}\label{eq3.2}
    n_a = \frac{1}{4 \pi t D}  \Gamma^{\frac{t}{\tau}} \exp \left( - \frac{r^2}{4 D t} \right) \Theta (t)
\end{equation}

 This solution describes the distribution of avalanches in an accelerating region that has no boundaries in the transverse plane. Otherwise, the solution must satisfy additional boundary conditions. Since electrons born outside the accelerating region do not cause new avalanche creation, it would be logical to consider the concentration of avalanches at the edge equal to zero. Therefore the dynamics of avalanches in acceleration region with finite transverse size might be described with the following equation with boundary and initial conditions

\begin{equation}\label{eq3.3}
    \begin{cases}
     \frac{\partial n_a}{\partial t} -  \triangledown \cdot (D \cdot \triangledown n_a) - \frac{n_a}{\tau^*} = 0\\
      n_a(t,r)|_{r=R} = 0\\
      n_a(t,r)|_{t=0} = n_I\\
    \end{cases}
\end{equation}

$n_I$ denotes the initial distribution of avalanches. The solution can be found in the following form:

\begin{equation}\label{eq3.4}
    n_a= \sum_{k = 1}^{\infty } T_k(t) X_k(r)
\end{equation}

Solving the problem on the coordinate-dependent part we can obtain that $X_k(r) = J_0 (\frac{r \mu_k}{R})$, where $J_0$ is the Bessel function and $\mu_k$ its zero. 

Substitution a series into the equation (\ref{eq3.3}) gives the equation for the time-dependent component

\begin{equation}\label{eq3.5}
    T_k(t) + T_k(t) \left( \frac{\mu_k^2 D}{R^2} - \frac{1}{\tau^*} \right) = 0
\end{equation}

The initial value of $T_k(t)$ can be found from the initial conditions on the avalanches distribution and the orthogonality of the Bessel functions:

\begin{equation}\label{eq3.6}
    T_k(0) = \frac{\int_0^R n_I(r) J_0(\frac{\mu_k r}{R}) r dr }{\int_0^R  J_{0}^2(\frac{\mu_k r}{R}) r dr}
\end{equation}

Let $A_k = T_k(0) \int_0^R  J_{0}(\frac{\mu_k r}{R}) dr$.  Then the total number of avalanches in the acceleration region at time $t$ is

\begin{equation}\label{eq3.7}
    N(t) =  \sum_{k = 1}^{\infty } A_k   \cdot \exp\left(- \left( \frac{\mu_k}{R} \right)^2 D t \right)  \exp \left(\frac{t}{\tau*}\right) 
\end{equation}

Assuming that $t=i \cdot \tau$, which corresponds to the time when $i$ generations of avalanches were born, the number of avalanches in $i$ generation is
\begin{equation}\label{eq3.8}
   N_i =  \sum_{k = 1}^{\infty } A_k   \cdot  \Gamma^i \alpha_k^i
\end{equation}

where $\alpha_k = \exp(- \left( \frac{\mu_k}{R} \right)^2 D \tau)$

\section{Correction to the criterion}\label{sec3}

The finite transverse size may lead to a decrease in the number of new avalanches and an increase in the requirements for self-sustaining RREA production by feedback. The obtained equation (\ref{eq3.8}) gives us the opportunity to find a correction to the criterion of self-sustaining RREA production. If $\Gamma \cdot \alpha_1$ is at least slightly more than 1 then other terms in the series decrease over time. Therefore it is enough to consider only the first term

\begin{equation}\label{eq4.1}
   N_i \approx A_1  \cdot  \Gamma^i \alpha_1^i = A_1  \cdot  \Gamma_{d}^i 
\end{equation}

where $\Gamma_d = \Gamma \cdot e^{-\left(\frac{2.405}{R}\right)^2 D \tau}$.

Thus, the criterion of self-sustaining production with correction is
\begin{equation}\label{eq4.1}
 \Gamma_d \geq 1
\end{equation}

\section{Geant4 simulation}\label{sec4}

The Monte-Carlo simulation was performed via Geant4 to obtain value of the diffusion coefficient, which describes how strongly RREA starting points propagate in the transverse direction due to relativistic feedback. In this tool a physical list can be chosen to determine the processes that need to be taken into account. In this work G4EmStandartPhysics option4 physics list was chosen, which contains all necessary processes, including Compton scattering, photoelectric effect and pair production for energies characteristic for RREA processes \cite{skeltved2014modeling, agostinelli2003geant4}. The simulation took place in a cylindrical volume, which was filled with air with a density corresponding to an altitude of 10 km above sea level, $\rho = 0.41 kg/m^3$. For the longitudinal size of the area and the field strength, the following values were chosen: $E = 300 kV/m$ and $L=445.7$ m. Such parameters according to \cite{stadnichuk2022criterion} provide a feedback coefficient $\Gamma = 1$. The easiest way is to get the diffusion coefficient from a distribution (\ref{eq3.2}). Therefore, a sufficiently large radius of the cylinder $R = 2000m$ was chosen. With such a radius, boundary conditions could be neglected, since avalanches launched from the center of the cylinder will not create new generations beyond the edge of the accelerating region.

The analytical consideration of diffusion does not take into account the coordinate of the beginning of the avalanche along the z-axis and describes only the two - dimensional distribution of avalanches in the transverse plane. Therefore, in order to correctly estimate the diffusion coefficient, it is necessary to consider an average avalanche. The simplest way to do this in simulation is to get the avalanche distribution and then launch avalanches with this distribution from the center of the cylinder. This is equivalent to launching average avalanches from the center. Thus, simulation was divided into four steps.

The first two steps are needed to obtain the distribution of the avalanches along the z-axis. First, the seed electrons were launched at the beginning of the electric field region. These electrons form RREAs, which radiate gamma-rays via bremsstrahlung. The energy, position, and momentum of the positrons generated by these gamma rays were recorded. After that, in the second step of the simulation, recorded positrons were launched. Electrons generated by these positrons were recorded. Thus, the obtained distribution is shown in Figure 1. It is worth noting that the form of the distribution is consistent with the distribution obtained analytically in the paper \cite{stadnichuk2022criterion}.

\begin{figure}
    \centering
    \includegraphics[width=1\linewidth]{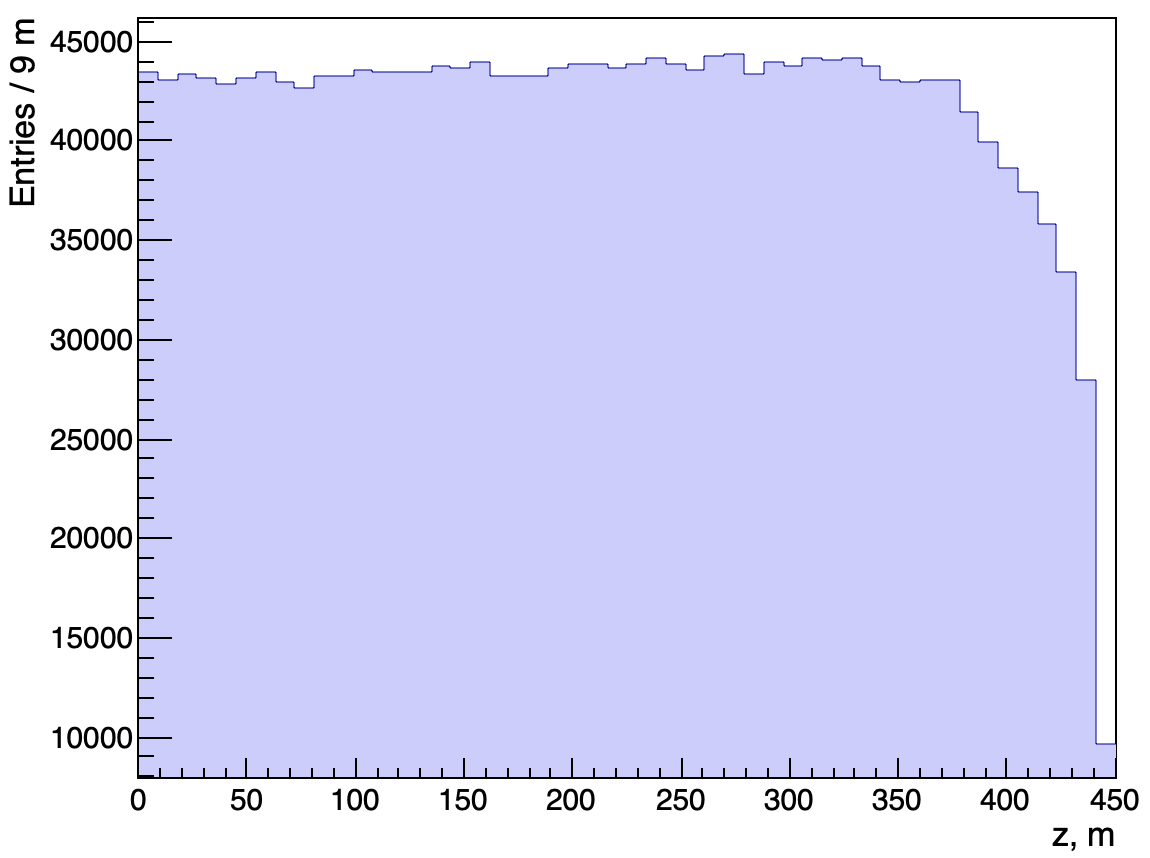}
    \caption{RREA starting point distribution, which was obtaind from electrons distribution along z-axis after second step of Monte-Carlo simulation. These electrons were created via feedback mechanism. It can be seen that shape of distribution is similar to analytically obtained one. \cite{stadnichuk2022criterion}.}
    \label{spectr_compare}
\end{figure}

In the third step obtained electrons were launched with recorded $z$ coordinate and $x = 0$, $y = 0$. Finally, the fourth step consisted of launching recorded positrons from the third step. Electrons generated by these positrons were recorded. The transverse distribution of these electrons is shown in Figure 2. It differs from the second-generation avalanche distribution by a constant factor $p_e$ - the probability that an electron turns around and runs away. This distribution was fitted according to formula (\ref{eq3.2}). The value of the diffusion coefficient multiplied by the time between generations was obtained from the fit: $D \tau \approx 836 \text{ }m^{2}$. This gives us the opportunity to evaluate the correction to the criterion of infinite feedback, therefore, the first four alpha coefficients are shown in Figure 3. As it was mentioned in section 4, $\alpha_1$ in the first term of the series corresponds to the correction to the feedback coefficient $\Gamma$. Moreover, all other alpha coefficients decrease with the growth of their number.

\begin{figure}
    \centering
    \includegraphics[width=1\linewidth]{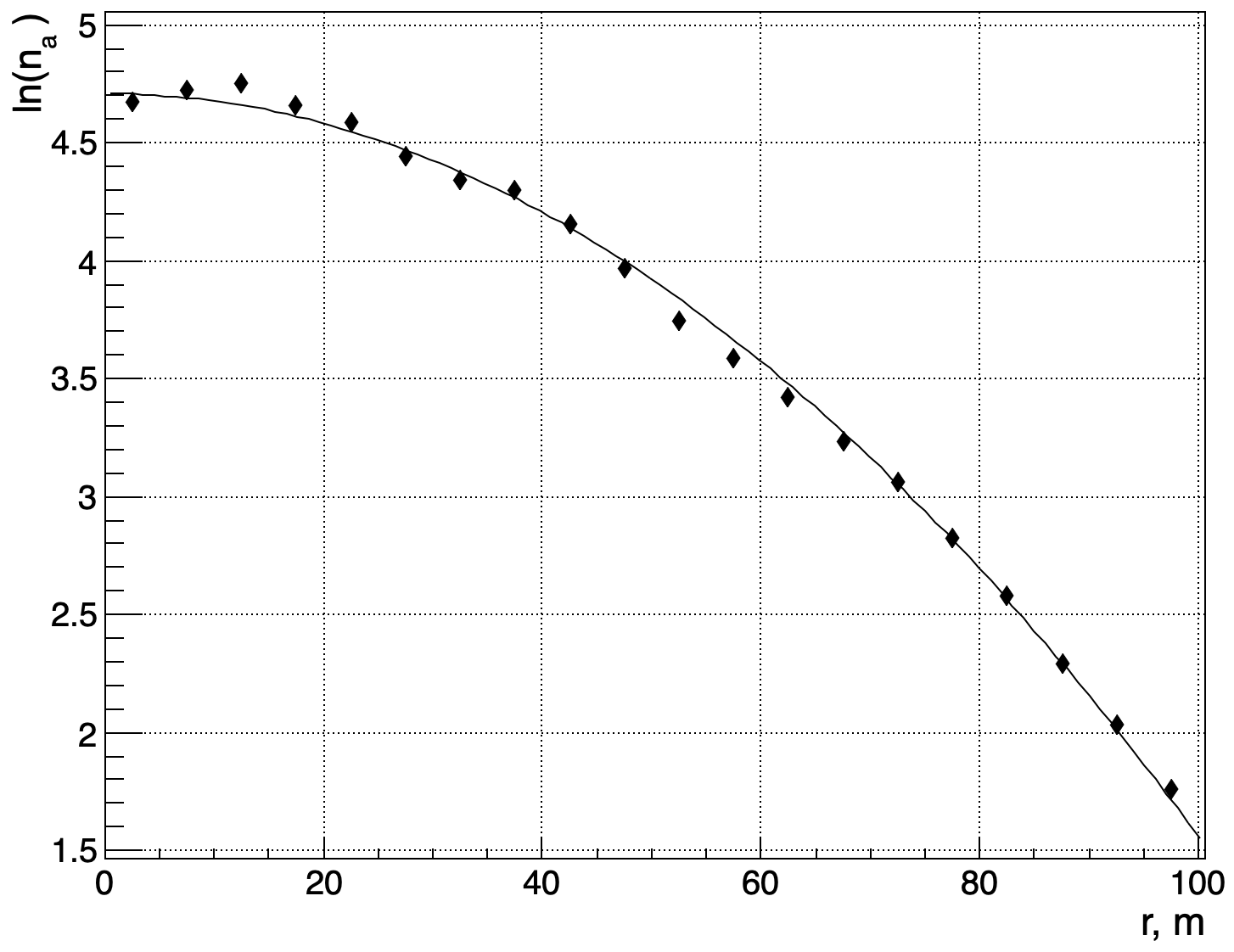}
    \caption{The logarithm of the transverse density of electrons recorded after fourth step of Monte-Carlo simulation. These electrons were created via feedback mechanism and they are the starting points of secondary avalanches. Diamonds - simulation result. Solid line - fit according to formula (\ref{eq3.2}). It can be seen from the Figure that the simulation results are in good agreement with the analytical model.}
    \label{spectr_compare}
\end{figure}

\begin{figure}
    \centering
    \includegraphics[width=1\linewidth]{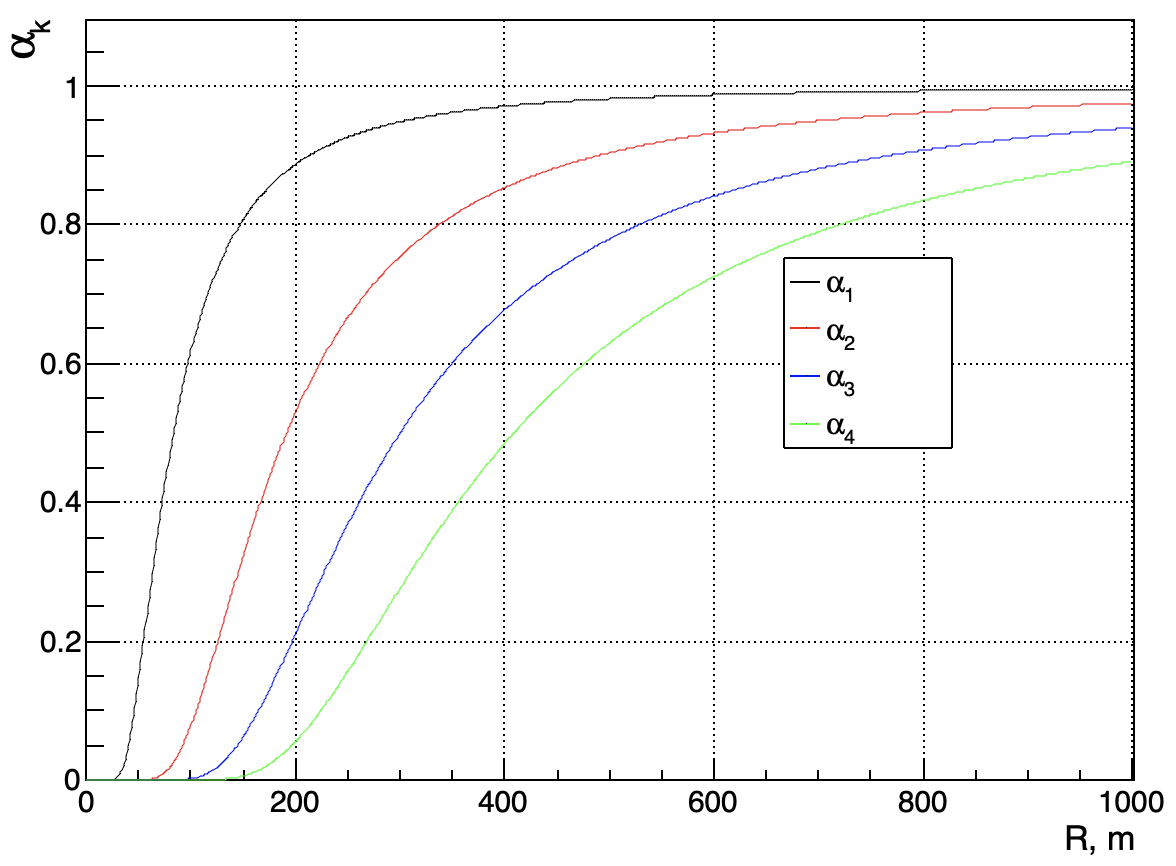}
    \caption{Dependence of the $\alpha_i$ from transverse size of accelerating region $R$ for the first four members of series (\ref{eq3.8}). $\alpha_{i}$ corresponds to correction factor to the feedback coefficient for each member of the series, which describes the number of avalanches. As it was said in Section 5, $\alpha_{1}$ gives the greatest contribution to the sum of the series. For the transverse size comparable with longitudinal one ($R \approx 223$ m) $\alpha_1 = 0.91$, $\alpha_2 = 0.60$}
    \label{spectr_compare}
\end{figure}

\begin{figure}
    \centering
    \includegraphics[width=1\linewidth]{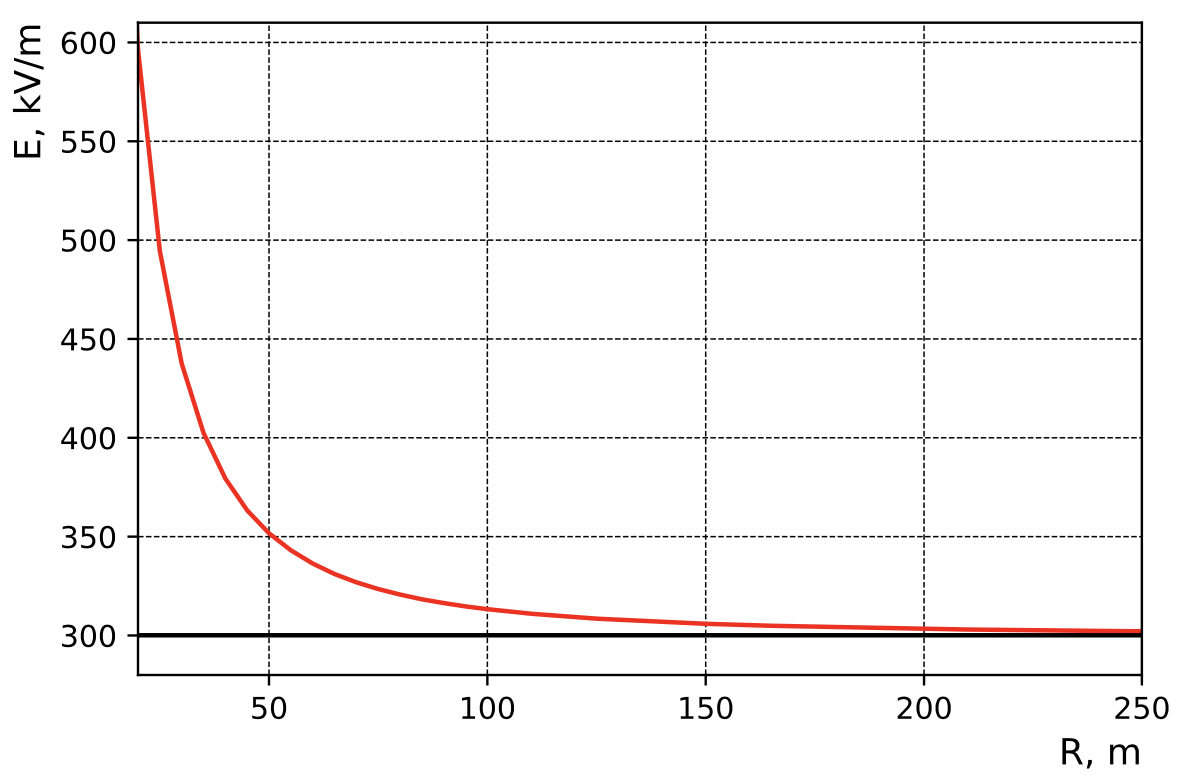}
    \caption{The dependence of the electric field strength at which an infinite feedback occurs on the transverse size of accelerating region (red line). The bold black line indicates the field strength necessary for positive feedback without taking into account the correction to the criterion. For the transverse size comparable with longitudinal one ($R \approx 223$ m) $E = 300.2$ kV/m.}
    \label{spectr_compare}
\end{figure}

\section{\label{disc}Discussion}

The expressions obtained in Sections 3 and 4 for the feedback coefficient allow us to calculate the field strength required for the occurrence of the infinite feedback regime. Moreover, these expressions allow us to search for the minimum transverse size of the accelerating region. It can be calculated from the corrected feedback coefficient $\Gamma_d$  and the rate of the TGF signal growth. This can be used, for example, to determine the size of regions with a uniform electric field. However, the presence of the diffusion coefficient and the average time $\tau$ in these expressions complicates the application of the expressions obtained. These parameters depend on the electric field strength and the length of the accelerating region and therefore must be calculated or measured. An attempt to solve this problem was made in Section 5, where the diffusion coefficient multiplied by time $\tau$ was evaluated by Monte Carlo simulation. Thus, we can use the obtained formulas only for a cell with a uniform electric field $E=300$ kV/m and a longitudinal size $L = 445.7$ m. 

The parameters of the accelerating region used in the simulation impose strong restrictions on the use of the coefficients obtained. However, as was done in \cite{stadnichuk2022criterion}, it can be assumed that the diffusion coefficient is determined only by the path length of the photon. This length, according to \cite{stadnichuk2021relativistic}, almost does not depend on the strength of the electric field. The average time between two avalanche generations is determined by the average velocity of electrons and positrons. In the work \cite{coleman2006propagation} it was shown that the average velocity changes slightly with the electric field strength, but remains close to 0.89c. Therefore, in the first assumption the value of $D \tau$ does not depend on the field strength. Using this assumption, for an accelerating region with a length of $445.7$ m, it is possible to obtain the dependence of the intensity, at which infinite feedback occurs, on the transverse size (Figure 4). For the transverse size comparable with longitudinal one ($R \approx 223$ m) $E = 300.2$ kV/m. However, this result was obtained only for an area with a certain longitudinal size L. Therefore,  study of the influence of the size of accelerating region on transversal diffusion is of great interest. Moreover, it is important to note that our model can only be applicable when the transverse size of the region is larger than the transverse size of one avalanche. Otherwise, the feedback can be significantly affected by the diffusion of electrons in an avalanche, which was studied in \cite{dwyer2010diffusion}.

\section{\label{disc}Conclusion}

The main purpose of this work was to describe the transverse propagation of avalanches as a result of feedback. This was done using a two-dimensional diffusion equation. From the solution of the equation, a сorrection to the minimal conditions for the self-sustaining feedback was obtained.

It was shown that the effect of the limited transverse size of the accelerating region on the feedback coefficient is small if the transverse size of the region is much larger than the longitudinal one. It becomes necessary to take diffusion into account when the transverse size becomes smaller than the longitudinal one. In this case, the correction to the electric field required for infinite feedback becomes extremely  significant. This result was obtained for the accelerating region with a longitudinal size $L = 445.7$ m. 

The aim of further research will be to analytically or via Monte-Carlo simulation obtain the dependence of the diffusion coefficient and the average time of new generation formation on the electric field strength and the longitudinal length of the accelerating region.

\newpage
\bibliography{biblio}

\end{document}